\documentclass[fleqn,twoside]{article}
\usepackage{multicol}
\usepackage[headings]{espcrc2}
\usepackage{psfrag}
\readRCS
$Id: espcrc2.tex,v 1.2 2004/02/24 11:22:11 spepping Exp $
\usepackage{epsfig}

\def\be{\begin{equation}}
\def\ee{\end{equation}}
\def\bea{\begin{eqnarray}}
\def\eea{\end{eqnarray}}

\def\eqref#1{eq.~(\ref{#1})}
\def\eqn#1{eq.~(\ref{#1})}

\def\sst{\scriptscriptstyle}
\def\hf{{\textstyle{1\over2}}}
\def\ihf{{\textstyle{i\over2}}}
\def\tr{\mathop{\rm tr}\nolimits}

\def\spa#1.#2{\left\langle#1\,#2\right\rangle}
\def\spb#1.#2{\left[#1\,#2\right]}

\def\Ab#1#2#3{\langle#1|#2|#3]}

\def\B#1#2{[#1#2]}

\def\Tri{{\rm Tri}}
\def\F{{\rm F}}

\def\Li{{\rm Li}}

\def\e{\epsilon}

\title{
One-loop helicity amplitudes for $H \to $ gluons: the all-minus configuration}

\author{S.D.\ Badger
\address[IPPP]{Institute of Particle Physics Phenomenology,
        University of Durham, Durham, DH1 3LE, UK}
and E.W.N.\ Glover
\addressmark[IPPP]
}

\begin{document}

\unitlength 1cm

\begin{abstract}
We use twistor inspired rules to compute the one-loop amplitude for 
a Higgs boson coupling to any number of negative helicity gluons in the
large top mass limit.
\end{abstract}

\maketitle

\thispagestyle{myheadings}
\markright{IPPP/06/47, DCPT/06/94, hep-ph/0607139}

\section{Introduction} 
Recent twistor inspired ideas have led to new (on-shell) ways of computing both tree and
loop graphs~\cite{Witten:twstr,CSW:MHVtree,BCF:rec,BCFW:proof,BCF:gu}.  
While there is still much work to be done in
determining the most efficient way of utilising these on-shell methods, it is clear that they
are extremely powerful in deriving analytic results for specific helicity
configurations of processes with an arbitrary number of gluons.
 
As an application of the on-shell methods, 
here we consider the one-loop amplitude for a Higgs boson coupling to
an arbitrary number of negative helicity gluons in the heavy top quark
approximation.

\section{The Higgs Model}

In the Standard Model the Higgs boson couples to gluons through
a fermion loop. The dominant contribution is from the top quark.
For large $m_t$, the top quark can be integrated out leading
to the effective interaction~\cite{HggOperator2,HggOperator5},
 \be
 {\cal L}_{\sst H}^{\rm int} =
  {C\over2}\, H \tr G_{\mu\nu}\, G^{\mu\nu}  \ .
 \label{HGGeff}
 \ee
In the Standard Model, and to leading order in $\alpha_s$,
the strength of the interaction is given by $C = \alpha_s/(6\pi v)$,
with $v = 246$~GeV. 

The MHV or twistor-space structure of the Higgs-plus-gluons amplitudes is best
elucidated \cite{DGK} by considering $H$ to be the real
part of a complex field $\phi = {1\over2}( H + i A )$, so that
\bea
 {\cal L}^{\rm int}_{H,A} &=&
{C\over2} \Bigl[ H \tr G_{\mu\nu}\, G^{\mu\nu}
             + i A \tr G_{\mu\nu}\, {}^*G^{\mu\nu} \Bigr]
 \label{effinta}\nonumber \\
\lefteqn{=
C \Bigl[ \phi \tr G_{{\sst SD}\,\mu\nu}\, G_{\sst SD}^{\mu\nu}
 + \phi^\dagger \tr G_{{\sst ASD}\,\mu\nu} \,G_{\sst ASD}^{\mu\nu} \Bigr]
 }\nonumber \\
 \label{effintb}
\eea
where the purely selfdual (SD) and purely anti-selfdual (ASD)
gluon field strengths are given by
$$
G_{\sst SD}^{\mu\nu} = \hf(G^{\mu\nu}+{}^*G^{\mu\nu}) \ , \quad
G_{\sst ASD}^{\mu\nu} = \hf(G^{\mu\nu}-{}^*G^{\mu\nu}) \ , 
$$
with
$$
{}^*G^{\mu\nu} \equiv \ihf \epsilon^{\mu\nu\rho\sigma} G_{\rho\sigma} \ .
$$
The important observation of \cite{DGK} was that, due to selfduality, the amplitudes for
$\phi$ plus $n$ gluons, and those for $\phi^\dagger$ plus $n$ gluons,
each have a simpler structure than the amplitudes for
$H$ plus $n$ gluons. Amplitudes can be constructed for
$\phi$ plus $n$ gluons  
and for $\phi^\dagger$ plus $n$ gluons separately.
Since $H = \phi + \phi^\dagger$,
the Higgs amplitudes can be recovered
as the sum of the $\phi$ and $\phi^\dagger$ amplitudes.

\section{Tree-level results}

The tree-level amplitudes can be decomposed into
color-ordered partial amplitudes as
\bea
\lefteqn{{\cal A}^{(0)}_n(\phi,\{k_i,\lambda_i,a_i\}) = 
i C g^{n-2}
\sum_{\sigma \in S_n/Z_n} }\nonumber \\
\phantom{~}\hspace{-3mm}&\times&\hspace{-3mm}
\tr(T^{a_{\sigma(1)}}\cdots T^{a_{\sigma(n)}})\,
A^{(0)}_n(\phi,\sigma(1^{\lambda_1},..,n^{\lambda_n})) 
\label{TreeColorDecomposition}
\eea
where $S_n/Z_n$ is the group of non-cyclic permutations on $n$
symbols, and $j^{\lambda_j}$ labels the momentum $k_j$ and helicity
$\lambda_j$ of the $j^{\rm th}$ gluon, which carries the adjoint
representation index $a_i$.  The $T^{a_i}$ are fundamental
representation SU$(N)$ color matrices, normalized so that
$\tr(T^a T^b) = \delta^{ab}$.  The strong coupling constant is
$\alpha_s=g^2/(4\pi)$.

The decomposition of the $HGG$ vertex into the selfdual and the anti-selfdual
terms \eqn{effintb}, guarantees that the whole class of (colour ordered)
tree-level helicity amplitudes must vanish
\cite{DGK};
\bea
 A^{(0)}_n(\phi,g_1^\pm,g_2^+,g_3^+,\ldots,g_n^+) &=& 0 \, , \\
  A^{(0)}_n(\phi^\dagger,g_1^\pm,g_2^-,g_3^-,\ldots,g_n^-) &=& 0 \, ,
 \label{phimpvanish}
\eea
for all $n$.

The tree amplitudes, with precisely two negative helicities
are the
first non-vanishing $\phi$ amplitudes. These amplitudes are
the $\phi$-MHV amplitudes.
General factorization properties now imply that they have to be extremely
simple \cite{DGK}, and when legs $q$ and $p$ have negative helicity,
are given by 
\bea
\lefteqn{A^{(0)}_n(\phi,g_1^+,g_2^+,\ldots, g_p^-, \ldots, g_q^-, \ldots ,g_n^+) }\nonumber\\
&=&
 { {\spa{p}.{q}}^4 \over \spa1.2 \spa2.3 \cdots \spa{n-1,}.{n} \spa{n}.1 } \,.
\label{eq:phi-mhv}
\eea
 In fact, the expressions
\eqn{eq:phi-mhv} for $\phi$-MHV $n$-gluon amplitudes are exactly the same
as the MHV $n$-gluon amplitudes in pure QCD. The only difference of
\eqn{eq:phi-mhv} with pure QCD is that the total momentum carried by gluons,
$p_1+p_2+\ldots+p_n=-p_{\phi}$ is the momentum
carried by the $\phi$-field and is non-zero.

The tree amplitude with all negative helicity gluons, the $\phi$-all-minus amplitude,
 also has a simple structure,
\bea
	\lefteqn{A^{(0)}_n(\phi;1^-,\ldots,n^-)} \nonumber\\
	&=& \left(-1\right)^n
	{m_H^4 \over \spb1.2 \spb2.3 \cdots \spb{n-1,}.{n} \spb{n}.1}\ .
	\label{eq:allminus}
\eea

Amplitudes with fewer (but more than two) negative helicities have been 
computed with Feynman diagrams (up to 4 partons) in Ref.~\cite{DFM}
and using MHV rules and recursion relations in Refs.~\cite{DGK,BGK}.
 
\section{The one-loop all-minus amplitude}

At one-loop  the colour ordered decomposition can be written, 
\bea
\lefteqn{{\cal A}^{(1)}_n(\phi,\{k_i,\lambda_i,a_i\}) = 
i C g^{n}
\sum_{\sigma \in S_n/Z_n} }\nonumber \\
&\times&\hspace{-3mm}N \tr(T^{a_{\sigma(1)}}\cdots T^{a_{\sigma(n)}})\,
A^{(1)}_n(\phi,\sigma(1^{\lambda_1},..,n^{\lambda_n}))\nonumber \\
&&{\rm + subleading~terms}.
\label{LoopColorDecomposition}
\eea
The terms subleading in the number of colours can be reconstructed
from the leading colour piece~\cite{BDDK:unitarity1}.

Expressions for the three-gluon amplitudes are available in
Ref.~\cite{Schmidt:1997wr} while numerical results for the 
helicity-summed four gluon amplitudes are given 
in~\cite{Ellis:2005qe}.

The one-loop amplitude is the sum of a cut-constructible piece and
a rational part,
\be
A^{(1)}_n = A^{(1),CC}_n + A^{(1),NCC}_n.
\ee
\subsection{The cut constructible contribution}

In a landmark paper, Brandhuber, Spence and Travaglini~\cite{Brandhuber:n4}  showed that it is
possible to calculate one-loop MHV amplitudes in ${\cal N}$=4~SYM using MHV rules.  
The calculation has many similarities to the unitarity based approach of 
Refs.~\cite{BDDK:unitarity1,BDDK:unitarity2},  the main difference being
that the MHV rules reproduce the cut-constructible parts of the amplitude directly, 
without having to worry about double counting.
This is the method that we wish to employ here.

The four-dimensional cut-constructible part of one-loop amplitudes can be constructed by joining two on-shell 
vertices by two scalar propagators, both of which need to be continued off-shell. 
We can exploit the tree-level vertices with an arbitrary number of gluons 
of Eqs.~\ref{eq:phi-mhv} and \ref{eq:allminus} to compute the one-loop amplitude with an arbitrary number of
negative helicity gluons.
	\label{fig:mhvdiags}
\begin{figure}[h]
	\psfrag{phi}{$\phi$}
	\psfrag{A}{$(a)$}
	\psfrag{B}{$(b)$}
	\psfrag{C}{$(c)$}
	\psfrag{p}{{\small $+$}}
	\psfrag{m}{{\small $-$}}
	\psfrag{1m}{$1^-$}
	\psfrag{2m}{$2^-$}
	\psfrag{nm}{$n^-$}
	\psfrag{jm}{$j^-$}
	\psfrag{jp1m}{$(j+1)^-$}
	\begin{center}
		\includegraphics[width=4cm]{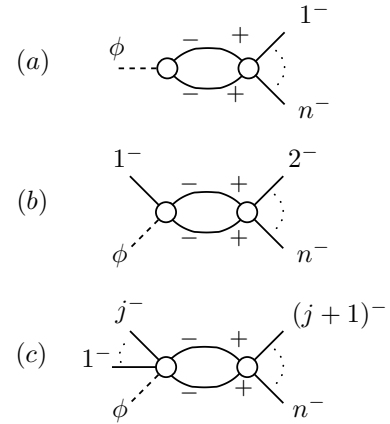}
	\end{center}
	\vspace{-10mm}
	\caption{The three classes of diagram contributing to the cut constructible part of
	the one-loop all-minus amplitude $A^{(1)}(\phi;1^-,\ldots,n^-)$.}
\end{figure}
There are three distinct contributions shown in Fig.~1.  Each class of diagrams
is composed of a phi-all-minus amplitude and a pure gluon MHV amplitude.
Each diagram represents an integral over the loop
momenta of the product of the two vertices. 
Note that only gluons circulate in the loop.   Fermions loops are eliminated by the specific 
helicity choice.  
The full amplitude is obtained by summing over the cyclic permutations.

The contribution  
from a particular graph is represented by
\bea
\lefteqn{	\frac{1}{(2\pi)^4}\int
	\frac{d^4L_1}{L_1^2}\frac{d^4L_2}{L_2^2}\delta^{(4)}(L_1+P-L_2)} \nonumber\\
	&&\times A_L(l_2^{-},\ldots,-l_1^{-})
	A_R(l_1^{+},\ldots,-l_2^{+}), 
	\label{eq:mhvint}
\eea
where
\be 
	L_i = l_i +z_i\eta,
\ee
and both $l_i$ and the reference vector $\eta$ are lightlike. 

If we consider the contribution from Fig~1(a), we find that
the product of tree amplitudes is:
\bea
	A_LA_R &=& \frac{m_H^4}{\B{l_1}{l_2}\B{l_2}{l_1}}
	\frac{(-1)^{n}\B{l_1}{l_2}^3}{\B{l_2}{1}\B{n}{l_1}\Pi^{n-1}_{\alpha=1} \B{\alpha}{\alpha+1}}\nonumber\\
	&=& A^{(0)}(\phi;1^-,\ldots,n^-)\frac{\B{l_1}{l_2}\B{1}{n}}{\B{l_2}{1}\B{n}{l_1}} \ .
	\label{eq:s1namps}
\eea 
Using momentum conservation this can be simplified such that the numerator is independent of $l_1$
and $l_2$. 
The delta function in \eqn{eq:mhvint} reads $\delta^{(4)}(L_1+P-L_2) = \delta^{(4)}(l_1+\widehat{P}-l_2)$,
such that,
\be P = p_1+\ldots+p_n
\ee
and
\be
	\widehat{P}   = P -(z_2-z_1)\eta\nonumber\\
			\equiv P -z\eta 
\ee
and applying $l_2=l_1+\widehat{P} $ together with some simple spinor algebra yields,
\bea
\frac{\B{l_1}{l_2}\B{1}{n}}{\B{l_2}{1}\B{n}{l_1}}  
&=& \frac{2\widehat{P}.1 \widehat{P}.n - \widehat{P}^2 1.n}{(\ell_1+1)^2(\ell_2+n)^2}\nonumber \\
&&-\frac{\widehat{P}.1}{(\ell_1+1)^2}
-\frac{\widehat{P}.n}{(\ell_2+n)^2} \ .\label{eq:integrand}
\eea
Following ref.~\cite{Brandhuber:n4} and simplifying the integration 
measure in \eqn{eq:mhvint}.
\bea
\lefteqn{	\frac{d^4L_1}{L_1^2}\frac{d^4L_2}{L_2^2} =
	\frac{dz_1}{z_1}\frac{dz_2}{z_2}d^4l_1d^4l_2\delta^{(+)}(l_1^2)\delta^{(+)}(l_2^2)}\nonumber\\
	\hspace{-5mm}
	&=&-2\frac{dzdz'}{(z-z')(z+z')}d^4l_1d^4l_2\delta^{(+)}(l_1^2)\delta^{(+)}(l_2^2),\nonumber \\
	\label{eq:intmeasure}
\eea
where $z=z_2-z_1$ and $z'=z_1+z_2$. At this point we notice that the integrand \eqn{eq:integrand}
is independent of the $z'$ variable and hence we can immediately integrate over this
variable leaving a four-dimensional phase space integral,
\be
-2\pi i \int\frac{dz}{z}d^4{\rm LIPS} 
\ee
where
\bea
\lefteqn{ d^4{\rm LIPS} = \frac{1}{(2\pi)^4}d^4l_1d^4l_2\delta^{(+)}(l_1^2)\delta^{(+)}(l_2^2)} \nonumber \\
\phantom{~}\hspace{-5mm} & \times&	 
	\delta^{(4)}(l_1-l_2+\widehat{P}) \ .
\eea

The remaining phase space integrals of the type shown in \eqn{eq:integrand}  can now be 
identified with the cuts of  
scalar box and triangle integrals.
The phase space integrals can be computed using the dimensional regularisation
schemeby promoting the phase space measure to $D=4-2\e$ dimensions
and have been evaluated by van Neerven~\cite{vanNeerven}.

The final integration over $z$ can be simplified by noting that
the result is independent of the choice of $\eta$ for each dispersion
integral~\cite{Brandhuber:n4}.
Special choices of 
$\eta$ such as $\eta=p_1$ or $p_n$
renders the integrand independent of $z$. 
The final dispersion integration over $z$ therefore reconstructs the 
part of the loop function which has the discontinuity as given by the
cut phase space integral.

\begin{figure}
	\psfrag{phi}{$\phi$}
	\psfrag{p}{{\small $+$}}
	\psfrag{m}{{\small $-$}}
	\psfrag{1m}{$1-$}
	\psfrag{2m}{$2-$}
	\psfrag{nm}{$n-$}
	\psfrag{jm}{$j-$}
	\psfrag{jp1m}{$(j+1)-$}
	\begin{center}
		\includegraphics[width=5cm]{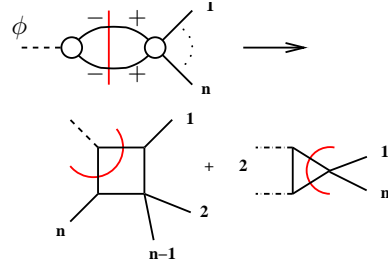}
	\end{center}
	\vspace{-10mm}
	\caption{Schematic of contribution from Fig 1(a).  
	}
	\label{fig:mhvdiags}
\end{figure}

As can be expected,  the contribution from Fig~1(a) is only associated with cuts in 
$s_{1,n} = m_H^2$.  Summing over the other diagrams and their permutations yields
the full unrenormalised cut-constructible result for the all-minus amplitude,
\begin{eqnarray}
\lefteqn{A_n^{(1),CC}(\phi,1^-,\ldots,n^-) = c_\Gamma A_n^{(0)}(\phi,1^-,\ldots,n^-)}\nonumber \\
&\times& \biggl (\sum_{i=1}^{n}\left(\Tri(s_{i,n+i-2})-\Tri(s_{i,n+i-1})\right)\nonumber\\
&&
\phantom{~}\hspace{-15mm}-\frac{1}{2}\sum_{i=1}^{n}\sum_{j=i+2}^{n+i-2}\F^{2me}(s_{i,j},s_{i+1,j+1};s_{i,j+1},s_{i+1,j})\nonumber \\
&&
\phantom{~}\hspace{-15mm}-\frac{1}{2}\sum_{i=1}^{n}\F^{1m}(s_{i,i+1},s_{i+1,i+2};s_{i,i+2})
\biggr ) 
\label{eq:cutresult}
\end{eqnarray}
where the dimensionally regularised loop functions are
\bea
	\Tri(s)= \frac{1}{\e^2}	 
	\left(\frac{\mu^2}{-s}\right)^{\e} \ ,  
\eea
\bea
	\lefteqn{F^{1m}(s,t;P^2)=}\nonumber\\
	\phantom{~}\hspace{-15mm}&& \frac{2}{\e^2}
	\bigg[
	\left(\frac{\mu^2}{-s}\right)^{\e}
	+\left(\frac{\mu^2}{-t}\right)^{\e}
	-\left(\frac{\mu^2}{-P^2}\right)^{\e}\bigg] \nonumber \\
	&&
	-2\Li_2(1-aP^2)
	-\frac{\pi^2}{3}
	\nonumber \\
	&&
	+2\Li_2(1-as)
	+2\Li_2(1-at) + {\cal O}(\e)
		\label{eq:f2me}
\eea
where $a=(P^2-s-t)/(-st)$ and
\bea
	\lefteqn{F^{2me}(s,t;P^2,Q^2)=}\nonumber \\
	&& \phantom{~}\hspace{-10mm}\frac{2}{\e^2}
	\bigg[
	\left(\frac{\mu^2}{-s}\right)^{\e}\hspace{-1mm}
	+\left(\frac{\mu^2}{-t}\right)^{\e}\hspace{-1mm}
	-\left(\frac{\mu^2}{-P^2}\right)^{\e}\hspace{-1mm}
	-\left(\frac{\mu^2}{-Q^2}\right)^{\e}\bigg] \nonumber \\
	&&
	-2\Li_2(1-bP^2)
	-2\Li_2(1-bQ^2)\nonumber \\
	&&
	+2\Li_2(1-bs)
	+2\Li_2(1-bt) + {\cal O}(\e)
		\label{eq:f2me}
\eea
where $b=(P^2+Q^2-s-t)/(P^2Q^2-st)$ and
\begin{equation}
	c_\Gamma = \frac{1}{(4\pi)^{2-\e}}\frac{\Gamma(1+\e)\Gamma^2(1-\e)}{\Gamma(1-2\e)}.
\end{equation}

For completeness, we note that
\be
A_n^{(1),CC}(\phi^\dagger,1^-,\ldots,n^-) =0,
\ee
so the full cut-constructible 
result for the Higgs coupling to any number of negative helicity gluons 
is given entirely by \eqn{eq:cutresult}.

\subsection{The non cut-constructible contribution}

The rational part of the amplitude cannot be reconstructed from
unitarity cuts.   We note that it is possible to fix these terms 
using the collinear limits~\cite{BDDK:unitarity1,BDDK:unitarity2}. We also note in passing
that there has been recent progress 
in using recursion relations to determine these
contributions~\cite{rational}.
Here, we use Feynman diagrams and observe that the quark loop
contribution fixes the rational part. This reduces the number of external
legs in the relevant Feynman diagrams by one.
The amplitude for three negative gluons is given in~\cite{Schmidt:1997wr}.
For four gluons,   we find that the unrenormalised
amplitude is,
\bea
\lefteqn{A_4^{(1),NCC}(H,1^-,2^-,3^-,4^-)} \nonumber\\
&=& 
\frac{N_p}{96\pi^2}\Biggl[
-\frac{s_{13}\Ab{4}{1+3}{2}^2}{s_{123}\spb{1}.{2}^2\spb{2}.{3}^2}
+\frac{\spa{3}.{4}^2}{\spb{1}.{2}^2}\nonumber \\
&&
+2\frac{\spa3.4\spa4.1}{\spb1.2 \spb2.3}
+\frac{s_{12}s_{34}+s_{123}s_{234}-s_{12}^2}{2\spb1.2\spb2.3\spb3.4\spb4.1}
\Biggr]\nonumber\\
&& + {\rm 3~cyclic~perms}
\label{eq:ncc4}
\eea
where $N_p=2(1-N_F/N)$.

\subsection{Checks of the result}
\subsubsection{Infrared poles}
The singular limit of the one-loop amplitude has 
a very prescribed form~\cite{catani}.
Inserting the explicit expressions for the loop integrals 
in \eqn{eq:cutresult} and keeping only the singular
terms, we find the expected result
\bea
\lefteqn{A_n^{(1)}(H,1^-,\ldots,n^-) = -\frac{c_\Gamma}{\e^2} \left[\sum_{i=1}^n 	 
	\left(\frac{\mu^2}{-s_{i,i+1}}\right)^{\e}\right]}\nonumber \\
&&\times~ 
	A_n^{(0)}(H,1^-,\ldots,n^-) + {\cal O}(1).
\eea

\subsubsection{Collinear limit}
In the collinear limit, the four gluon amplitude factorises as illustrated in Fig.~3
onto a $H \to ggg$ amplitude multiplied by a $g \to gg$ splitting function.
We have checked that \eqn{eq:cutresult} and \eqn{eq:ncc4} together 
produce the correct limit~\cite{BDDK:unitarity1,BDDK:unitarity2}.
\begin{figure}
	\psfrag{phi}{$H$}
	\psfrag{pm}{{\small $\pm$}}
	\psfrag{mp}{{\small $\mp$}}
	\psfrag{p}{{\small $+$}}
	\psfrag{m}{{\small $-$}}
	\psfrag{1m}{$1-$}
	\psfrag{2m}{$2-$}
	\psfrag{nm}{$n-$}
	\psfrag{jm}{$j-$}
	\psfrag{jp1m}{$(j+1)-$}
	\begin{center}
		\includegraphics[width=6cm]{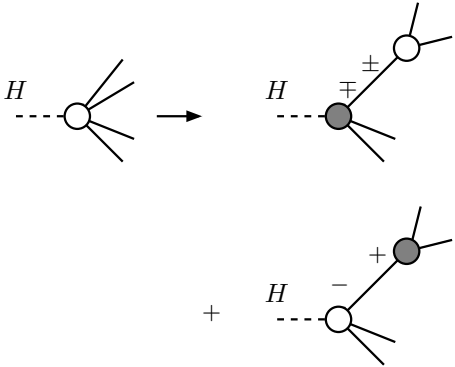}
	\end{center}
	\vspace{-10mm}
	\caption{Schematic of the collinear limit of the all-minus one-loop amplitude.
	Open circles represent loops, while shaded circles represent trees.
	}
	\label{fig:mhvdiags}
\end{figure}

\subsubsection{Soft Higgs limit}
For the case of a massless Higgs boson, we can consider
the kinematic limit $p_H \to 0$.
In this limit, because of the form of the $H G_{\mu\nu}G^{\mu\nu}$
interaction, the Higgs boson behaves like a constant,
so the Higgs-plus-$n$-gluon 
amplitudes should be related to
pure-gauge-theory  amplitudes~\cite{Dixon}.
In this limit, the tree-level all minus amplitude vanishes, and therefore so does
the cut-constructible contribution.
The rational part for four gluons \eqn{eq:ncc4}
simplifies considerably, the first and fourth terms in the square bracket 
vanish, while the other two terms combine using spinor
algebra and momentum conservation and we find,
\bea
\lefteqn{A_n^{(1)}(H,1^-,2^-,3^-,4^-)\phantom{\qquad\qquad\qquad\qquad~}} \nonumber \\
\phantom{~}\qquad\qquad&\longrightarrow& 
4 A_n^{(1)}(1^-,2^-,3^-,4^-)
\eea
where~\cite{Bern:1991aq},
\be
A_4^{(1)}(1^-,2^-,3^-,4^-)=
-\frac{N_p}{96\pi^2}\frac{\spa1.2\spa3.4}{\spb1.2 \spb3.4}.
\ee

\section{Outlook}
In this paper, we described the calculation of the cut-constructible part of
the one-loop amplitude
for a Higgs boson coupling to any number of negative helicity gluons in the
large top mass limit.  The finite rational contribution was presented for 
four gluons.  In principle, the on-shell methods
can be employed to compute both the rational terms for an arbitrary number of gluons
as well as the amplitudes for other helicity configurations.

\section{Acknowledgements}
EWNG gratefully acknowledges the support of PPARC through a Senior Fellowship
and SDB acknowledges the award of a PPARC studentship.   We thank Carola Berger,
Vittorio Del Duca and Lance Dixon for useful discussions and for pointing out a
number of misprints in the original version of this paper.


\end{document}